\documentclass[dvips]{elsart}
\usepackage{graphicx}    
\usepackage{amsmath}     

\begin{document}

\journal{Nucl. Instrum. Methods A}

\begin{frontmatter}

\title{
 Measurements of muon flux in the Pyh\"asalmi underground laboratory
}

\author[cuppp]{T.~Enqvist\corauthref{tc}}, \ead{timo.enqvist@oulu.fi}
\author[cuppo]{A.~Mattila},
\author[cuppp]{V.~F\"ohr},
\author[sgo]{T.~J\"ams\'en},
\author[cuppo]{M.~Lehtola},
\author[cuppo]{J.~Narkilahti},
\author[cuppo]{J.~Joutsenvaara},
\author[cuppo]{S.~Nurmenniemi},
\author[cuppo]{J.~Peltoniemi},
\author[cuppo]{H.~Remes},
\author[cuppo]{J.~Sarkamo},
\author[cuppp]{C.~Shen},
\author[sgo]{I.~Usoskin}

\address[cuppp]{CUPP, P.O. Box 22, FIN-86801 Pyh\"asalmi, Finland}
\address[cuppo]{CUPP. P.O. Box 3000, FIN-90014 University of Oulu, Finland}
\address[sgo]{SGO, P.O. Box 3000, FIN-90014 University of Oulu, Finland}

\corauth[tc]{Corresponding author.}

\begin{keyword}
 underground laboratory, measured muon flux, muon background, cosmic-ray
 induced muon, plastic scintillation detector
  \PACS  96.40.Tv\sep  
         95.55.Vj\sep  
         95.85.Ry\sep  
         29.40.Mc      
\end{keyword}

\begin{abstract} 

The cosmic-ray induced muon flux was measured at several depths in the 
Pyh\"a\-salmi mine (Finland) using a plastic scintillator telescope
mounted on a trailer.  The flux was determined at four different depths 
underground at 400 m (980 m.w.e), at 660 m (1900 m.w.e), at 990 m (2810 
m.w.e) and at 1390 m (3960 m.w.e) with the trailer, and also at the ground 
surface. In addition, previously measured fluxes from depths of 90 m (210 
m.w.e) and 210 m (420 m.w.e) are shown. A relation was obtained for the 
underground muon flux as a function of the depth. The measured flux follows 
well the general behaviour and is consistent with results determined in other 
underground laboratories.

\end{abstract}

\end{frontmatter}

\section{Introduction} 

Existing and future experiments on searching for proton decay or dark
matter, or on low-energy neutrino detection need a deep underground
location and low-background environment. The cosmic-ray muon flux can
be reduced several order of magnitudes by shielding experiments with
rock overburden.

Muons, however, are usually not the dominant background in underground
experiments as they can be identified and distinguished by relatively
simple way. Much more problematic contribution on the background comes
from high-energy neutrons generated by muons in the interactions with
the surrounding rock or with the detector material. The energy of these
neutrons may even extend up to the GeV region. This is, for example, the
factor limiting the current experiments on dark matter searches.

Precise knowledge of the muon flux is thus important in order to estimate, 
or calculate by Monte Carlo codes, the absolute underground neutron flux. 
In the present work, the underground cosmic-ray muon flux was systematically 
investigated by measuring it at several depths in the Pyh\"asalmi mine.

The Pyh\"asalmi mine (owned by Inmet Mining Corporation, Canada) is an 
active zinc, copper and pyrite mine in central Finland. It is the oldest 
and the deepest operational base-metal mine in Europe, extending down to 
about 1440 metres (corresponding about 4100 m.w.e). The site provides 
excellent opportunities for the research of underground physics by offering 
very stable bedrock, modern infrastructure, and good traffic conditions all 
around a year. CUPP (Centre for Underground Physics in Pyh\"asalmi) is 
establishing an international underground laboratory into the connection 
with the mine.

\section{Experimental details} 

The experimental setup consisted in total of 12 plastic scintillation
counters of type NE102A, having maximum light output at wavelength of 423 
nm. The size of a scintillation plate is 50 cm $\times$ 50 cm $\times$ 5
cm. Each counter was attached with Hamamatsu head-on type R329-02 
photomultiplier tube (PMT) using fishtail-shaped acrylic light quide
and BC-630 as an optical connector. The sensitivity range of the PMT is 
300 -- 600 nm.

Each scintillation plate and PMT was mounted on a water- and light-tight
stainless-steel box with size of 1.55 m $\times$ 0.65 m $\times$ 0.15 m, 
and the total weight of about 60 kg. The boxes were arranged in a geometry
covering a sensitive area of 1.5 m$^2$, and were installed on a movable 
trailer together with the data acquisition system (DAQ). The vertical 
distance (of the mid-planes) of two scintillation plates was 15 cm.

Two boxes were overlaid, in order to get a coincidence signals and reduce
noise signals. The DAQ consisted of two separate and identical systems, 
each one connected into six scintillators. Standard NIM electronics with 
discriminators and coincidence units were used together with 8-bit 
ADC~\&~Scaler units. UPS system was used to ensure secure data collection 
in the case of (probable) electrical break-down.

The scintillators were tested before the installation on the trailer. The
efficiency measurement was done by piling-up three detectors (stainless-steel
boxes), and the one to be tested was placed in the middle. The trigger was
taken from a coincidence unit which verified that muon had passed all the
counters. The efficiency of (97 $\pm$ 3)\% was obtained for each single 
detector.

\begin{table}
 \begin{center} \leavevmode
 \begin{tabular}{crcrr} 
 \hline\hline
     Depth     & Duration & Coordinates & Counts  & Flux Density       \\
 $[$m] ([m.w.e]) & [hours]  &   (x,y)   &         & [m$^{-2}$s$^{-1}$] \\
 \hline
     0         &     12   & (8308,2680) & 7600000 &  180 $\pm$ 20      \\ 
   400 (980)   &    192   & (8535,2635) &   13232 &                    \\
   400 (980)   &    132   & (8580,2657) &   10510 & 
                                    (2.1 $\pm$ 0.2) $\times$ 10$^{-2}$ \\
   660 (1900)  &    325   & (8390,2425) &    3850 & 
                                    (3.2 $\pm$ 0.3) $\times$ 10$^{-3}$ \\
   990 (2810)  &   1368   & (8400,2392) &    3282 & 
                                    (6.2 $\pm$ 0.6) $\times$ 10$^{-4}$ \\
  1390 (3960)  &   2748   & (8225,2460) &    1206 & 
                                    (1.1 $\pm$ 0.1) $\times$ 10$^{-4}$ \\
 \hline\hline
 \end{tabular} 
 \caption{Parameters of the measurements of the present work. In the first
  column the nominal depth and corresponding equivalent depth are given. 
  Coordinates are those used by the mine and they describe the position of 
  the trailer. The 'Counts' means the uncorrected number of muons collected.
  In the last column the deduced muon flux of the corresponding level is 
  shown. The average is given for the depth of 400 metres.
 }
 \label{table_params} 
 \end{center}
\end{table}

\section{Measurements} 

The muon-flux measurements were performed with the trailer at the ground 
surface and at 4 different depths underground, which were two positions at 
400 metres (corresponding 980 m.w.e), 660 metres (1900 m.w.e), 990 metres 
(2810 m.w.e), and 1390 metres (3960 m.w.e). In addition, data are shown for 
the depths of 90 and 210 metres which have been measured earlier with similar 
(but stationary) detector telescope. Their corresponding depths are 210 
m.w.e and 420 m.w.e, respectively.

The measurement durations were chosen such that the statistical uncertainty 
would be small (about 2\% or less). The measurement took about two weeks 
in 400 and 660 levels each, about two months at 990 levels, and about four 
months at 1390 level.
Details of the measurements are shown in Table \ref{table_params}.

\begin{figure}
 \begin{center}
  \leavevmode
  \includegraphics[scale=0.5,angle=-90.0]{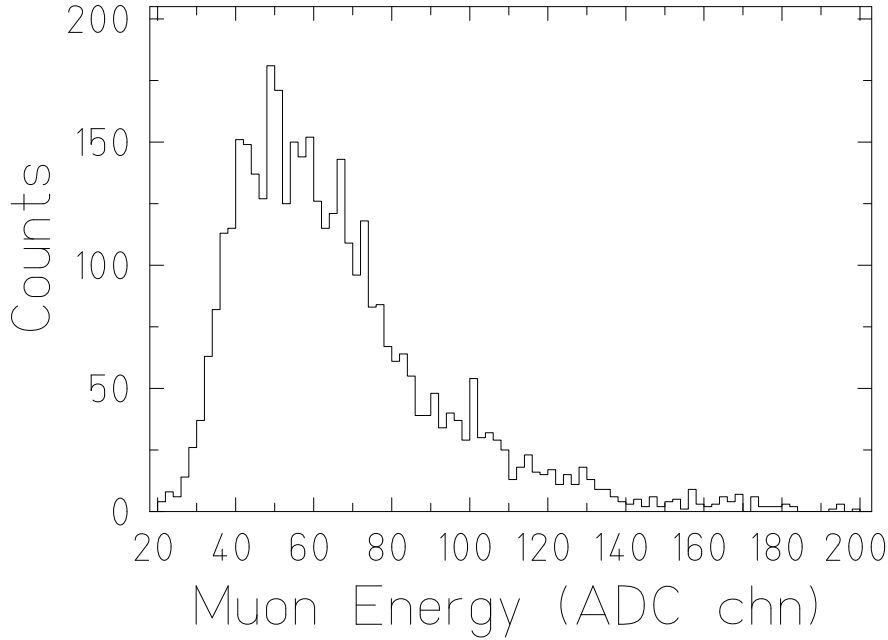}
  \caption{
   The total energy-loss spectra, in ADC-channels, collected at 660 metres 
   underground. The bin size is two channels. 
  }
  \label{figu_Edist} 
 \end{center}
\end{figure}

\section{Data analysis and results} 

The composition of the rock in the mine is well known, and the average 
rock density of 2.85 g/cm$^3$ was used to convert the depths in the 
m.w.e-units (metre water equivalent). The muon flux was obtained by 
dividing the measured counts by the measurement time (see Table 
\ref{table_params}) and the detector area (1.5 m$^2$). In the analysis 
vertical muons were not distinguished from non-vertical muons. 

The measurement positions of underground levels of 210 and 400 metres 
situate directly under the open pit mine which is nearly 100 metres depth, 
and contains also some loose rocks and sand. The metre-water-equivalent 
values of these two depth were taken as effective depth determined by 
simulation \cite{Kul03}. Their depth (as m.w.e) corresponds to flat surface 
geometry giving the same reduction in the flux.

The open pit mine has no (significant) effect on other measurement positions 
since they are not directly below it, and mostly lie deeper. For these
levels the metre-water-equivalent depth was determined from the vertical 
distance to the surface, taking into account possible underground caverns.

The number of muons were obtained from the energy-loss spectrum by counting
the events. The statistical uncertainty was taken as square root of the 
number of events. Fig. \ref{figu_Edist} shows, as an example, the total 
muon energy-loss spectrum measured at 660 metres underground.

The collected number of muons were corrected for the detector efficiencies
(being about (97\%)$^2$), and by the amount of muon events below the energy 
threshold which was estimated in the test measurement to be (5 $\pm$ 2)\%
\cite{Jam05}. Signals from neutrons induced by the rock activity or muons 
were not corrected for.  

The geometrical uncertainty, i.e. that the two scintillation plates were 
not fully overlapping, was estimated as 4\%. The uncertainty in the depth 
was taken as about 5 metres corresponding to about 20 m.w.e including also 
the variation in the average rock density.

The detector acceptance was taken into account by determining the effective
area of the scintillation pair at each depth using two different zenith
angle distribution (Miyake model and Intermediate-depth model) of cosmic-ray
muons. The distributions were taken from Ref. \cite{Gri01}, and they resulted 
the same effective area for 210-level and deeper. For the surface and 90-level 
the Miyake distribution was used. The acceptance correction decreases for 
deeper depths; for the surface the acceptance was corrected by a factor of 
1.42 and for the 1390-level by a factor of 1.23.  The uncertainty of the
correction was assumed to be better than 10\%.

\begin{figure}
 \begin{center}
  \leavevmode
  \includegraphics[scale=0.65]{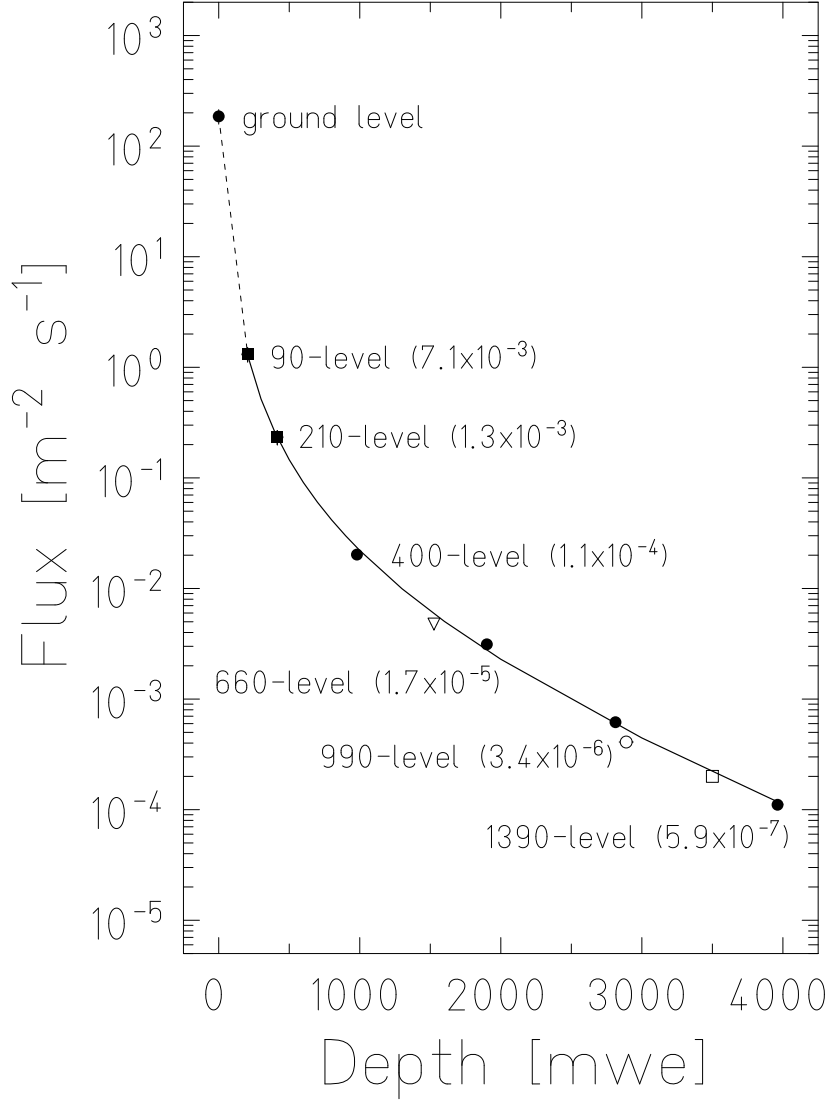}
  \caption{
   Full symbols present the fluxes measured in the Pyh\"asalmi mine:
   full circles are the result of the present work, full squares have 
   been measured earlier with similar detector system.
   In parentheses the corresponding reduction of the flux compared with
   the flux at the surface is given. 
   Open symbols are shown in comparison with other measurements: 
   open circle from Ref. \cite{Rob03},    
   open square from Ref. \cite{Arn03},    
   open triangle from Ref. \cite{Esc05}.  
   If not shown, the uncertainty is smaller than the size of a symbol.
   The solid line is the result of the fit of Eq. (\ref{eq_flux}).
  }
  \label{figu_vuot} 
 \end{center}
\end{figure}

The results of the muon fluxes measured in the present work are shown in 
Fig.~\ref{figu_vuot} as full circles. The flux at 400 level is an average of 
the two measurement. In addition, Fig. \ref{figu_vuot} shows fluxes from 
90- and 210-levels, which have been measured earlier with similar but 
stationary detector system \cite{Jam05}, as full squares. The reduction 
factor of the muon flux compared with the surface flux is shown in 
parentheses for each level. Open symbols show the results of other 
muon-flux measurements in some underground laboratories for comparison.

The behaviour of the total muon flux underground as a function of the 
depth was extracted from the data using six data points (full symbols in 
Fig. \ref{figu_vuot} except the surface flux) by fitting a function of the 
form \cite{For04}
\begin{equation}
 F(x) = A \cdot ( x_0 / x ) ^{\eta} 
          \cdot \exp( -x / x_0 ), 
 \label{eq_flux}
\end{equation}
where $F$ is the flux as m$^{-2}$s$^{-1}$ and $x$ the depth as m.w.e. 
The following values were obtained for the fit parameters (with
$\chi^2 \approx 0.007$):
\begin{align} 
      A & = (0.025 \pm 0.004) \  \rm{m}^{-2} \rm{s}^{-1}, \notag \\
   \eta & = 2.18 \pm 0.12,  \notag \\
    x_0 & = (1330 \pm 140) \ \rm{mwe}.  \notag
\end{align}
The fit function (Eq. \ref{eq_flux}) is plotted in Fig. \ref{figu_vuot}
with the above parametres as the solid line.

\section{Discussion} 

The flux measured in the present work is consistent with results obtained
in other underground laboratories at specific depths (Fig. \ref{figu_vuot}). 
The data agrees very well with the measurements done at Gran Sasso and WIPP
(Refs. \cite{Arn03,Esc05}), and reasonably well also with the result measured
at Boulby \cite{Rob03}. The deviation with the Boulby value at this specific 
depth is some 30\%. 

The best fit plotted in Fig. \ref{figu_vuot} as the solid line describes 
well the decrease of the flux as a function of the depth, and it can be used
to estimate the muon flux at an arbitrary depth with a good accuracy. The 
values of $\eta$ and $x_0$ are also relatively close to the values of other 
similar fit-results obtained for the behaviour of the vertical muon intensity 
as a function of the depth 
($\eta = 1.93^{+0.20}_{-0.12}, x_0 = (1155^{+60}_{-30})$ m.w.e) \cite{For04}.

\section{Conclusion} 

Cosmic-ray induced muon flux has been systematically measured at several
depths in the Pyh\"asalmi mine. The data appear to be the most comprehensive 
one of this type, and at specific points consistent with results determined 
in other underground laboratories. A relation obtained for the underground
muon flux as a function on the depth is also consistent with previous
investigations. However, since the measurements were done with the same
detector at the different depths, the systematical uncertainty between the
depths should be negligible. The muon-flux measurements will continue.

The fast neutron background is going to be measured in the Pyh\"asalmi 
underground laboratory during the next two years. The measured muon flux of
the present work offers already a good way to estimate the magnitude
of the muon-produced neutron flux. The results of the present work are 
also valuable in evaluating the suitability of the Pyh\"asalmi site as a 
underground laboratory hosting future experiments requiring effective 
cosmic-ray shielding.

\textbf{Acknowledgements}

The authors want to acknowledge the support and help of the local staff of 
the Pyh\"asalmi mine. 
The Department of Physics of the University of Jyv\"askyl\"a is also 
acknowledged by their contribution.
The Pyh\"asalmi underground laboratory is funded by the European Union 
Regional Development Fund (ERDF).

\end{document}